\documentclass{IEEEoj-data}
\usepackage{cite}
\usepackage{amsmath,amssymb,amsfonts}
\usepackage{algorithmic}
\usepackage{dirtree}
\usepackage{graphicx,color}
\usepackage{textcomp}
\usepackage{hyperref}
\usepackage{balance}
\usepackage{booktabs}
\usepackage{orcidlink}
\usepackage{siunitx}
\usepackage[capitalize]{cleveref}
\def\BibTeX{{\rm B\kern-.05em{\sc i\kern-.025em b}\kern-.08em
    T\kern-.1667em\lower.7ex\hbox{E}\kern-.125emX}}
\AtBeginDocument{\definecolor{ojcolor}{cmyk}{0.93,0.59,0.15,0.02}}

\makeatletter
\renewcommand{\ieeelogo}{\raisebox{3pt}{\normalfont\large\bfseries Preprint}}
\AtBeginDocument{%
  \def\ps@plain{%
    \def\@oddhead{\vbox{\hsize\textwidth\vspace*{-13pt}\vbox to 29pt{\hsize126pt\vspace*{5pt}\ieeelogo}\hfill\par\vspace*{-2.75pt}\hbox to \textwidth{\vrule width\textwidth height.3pt depth0pt}}}%
    \let\@evenhead\@oddhead%
    \def\@oddfoot{\hbox to \textwidth{{\rffont\thepage}\hfill{\rffont Preprint -- \today}}}%
    \def\@evenfoot{\hbox to \textwidth{{\rffont Preprint -- \today}\hfill{\rffont\thepage}}}%
  }%
  \pagestyle{plain}%
}
\makeatother

\begin{document}

\title{{\textcolor{black}{Descriptor:\\}}  \textit{A Hybrid Indoor and Indoor-Outdoor Positioning \\ Multi-Technology Dataset (HYMN)}}

\author{Muhammad Ammad\orcidlink{0009-0007-2720-8433}\authorrefmark{}, Albrecht Michler\orcidlink{0000-0002-3434-3488}\authorrefmark{}, Paul Schwarzbach\orcidlink{0000-0002-1091-782X}, Jonas Ninnemann\orcidlink{0000-0001-7988-079X}, Hagen Ußler\orcidlink{0009-0001-2212-6590}, AND Oliver Michler\orcidlink{0000-0002-8599-5304}\authorrefmark{}
}
\affil{Institute of Traffic Telematics, TUD Dresden University of Technology, 01069 Dresden}
\corresp{CORRESPONDING AUTHOR: Muhammad Ammad (e-mail: muhammad.ammad@tu-dresden.de).}

\begin{abstract}
This article introduces the \textbf{HYMN} (\textbf{HY}brid \textbf{M}ulti-technology \textbf{N}avigation) dataset: a multi-system, and time synchronized dataset for localization research based on opportunistic signals collected in an indoor-outdoor scenario. HYMN comprises measurement data collected in an industrial hall setting for five different positioning systems including Ultra-Wideband (UWB), Bluetooth Low Energy (BLE), WiFi, 5G, and Global Navigation Satellite System (GNSS). Unlike existing datasets that focus on single technologies or purely indoor/outdoor scenarios, HYMN combines five positioning technologies with explicit coverage of indoor-outdoor transitions, enabling multi-sensor fusion research for seamless localization. Each instance of data is identified through a unique measurement id and it represents time-stamped observations relevant for each system respectively along with the ground truth information. HYMN is designed to support a wide range of localization tasks including multi-sensor fingerprinting, cross-technology fusion, and seamless indoor-outdoor positioning. The synchronized measurements from GNSS and other terrestrial systems enable researchers to investigate how heterogeneous signals complement each other to overcome individual technology limitations such as GNSS degradation in covered areas or terrestrial system variability in dynamic environments.
 \\

{\textcolor{ieeedata}{\abstractheadfont\bfseries{DATA DOI/PID}}}     \href{https://doi.org/10.5281/zenodo.17979434}{DOI: 10.5281/zenodo.17979434}
  \\
 {\textcolor{ieeedata}{\abstractheadfont\bfseries{DATA TYPE/LOCATION}}} Ranging data; industrial hall; Germany
\end{abstract}

\begin{IEEEkeywords}
HYMN dataset, Radio localization dataset, signals of opportunity, seamless positioning, Ultra-Wideband (UWB), Bluetooth Low Energy (BLE), WiFi, 5G New Radio (NR), GNSS
\end{IEEEkeywords}

\maketitle

\section*{BACKGROUND}

Localization has gained significant attention in recent years, however the enabling radio systems are confronted with heterogenous challenges in different scenarios. Outdoor environments rely primarily on GNSS for localization which performs exceptionally well in open skies but its functionality is limited in covered areas for indoor positioning. The indoor-outdoor positioning requires seamless integration of the two technologies used for both the indoor and outdoor environments. Outdoor to indoor transitioning while positioning poses a lot of difficulties since it requires precise switching between the technology being used outside (e.g., GNSS) and inside (e.g, Wi-Fi, 5G, etc.) for positioning \cite{Zhu2019}. Accurate positioning within these transitional areas, which bridges indoor and outdoor environments, are often difficult to obtain with standard positioning models. Developing navigation algorithms involving such transitional areas requires extensive sensor readings  \cite{Wang2023}.

Studies show that conventional GNSS solutions achieve reliable outdoor accuracy but their performance deteriorates in obstructed or indoor areas due to multipath reflections, non line-of-sight (NLOS) interference, and signal blockage, thereby motivating the use of additional sensing modalities for continuous and resilient localization \cite{Cao2022, Wang2022}. Terrestrial and short-range positioning technologies such as WiFi, UWB, and BLE offer improved coverage and continuity in indoor or partially GNSS-denied environments. However, their performance is constrained by environmental dynamics and signal variability, motivating continued research into sensor fusion, and data-driven adaptation across heterogeneous settings \cite{Leitch2023, Li2018}. Complementary datasets and benchmarks that span multiple/mixed environments are critical for evaluating observability-aware estimators, multipath/NLOS resilience, and cross-system fusion strategies, thereby advancing robust localization under dynamic environmental conditions, sensor failures, and diverse deployment scenarios \cite{ElMowafy2018, Zhu2023}. However, a systematic review of 119 indoor positioning datasets reveals that 50\% lack information on collection area size, 21\% omit ground truth protocol documentation, and 13\% have undefined environment types, indicating that critical metadata required for reproducible benchmarking remains widely absent \cite{ordip}. Collectively, these studies highlight the necessity of multi-modal dataset, and fusion methodologies to ensure reliable performance across environments.

\begin{table*}[ht]
\centering
\caption{Overview of sensor systems and their identifiers}
\begin{tabular}{lllp{5.2cm}c}
\toprule
\textbf{System} & \textbf{System ID} & \textbf{Sensor/Hardware} & \textbf{Measurement Principle} & \textbf{Anchors} \\
\midrule
GNSS            & \texttt{gnss}      & NovAtel PwrPak 7                  & Pseudorange                  & varying      \\
UWB             & \texttt{uwb}       & ZigPos RTLS                       & Two-Way Ranging (TWR) / RTT                 & 10      \\
Bluetooth LE             & \texttt{ble}       & Metirionic DMK-215                & Phase-Based Ranging (PBR)                  & 5       \\
Wi-Fi           & \texttt{wifi}      & HPE Aruba 630 Series APs         & Fine Timing Measurement (FTM, RTT)         & 6       \\
5G NR           & \texttt{nr5g}      & Industrial Small Cell System     & Fingerprinting (SNR, TDOA-based)           & 3       \\
Ground Truth    & \texttt{ref}       & Leica TS16 Total Station         & Optical survey (total station)             & 1       \\
\bottomrule
\end{tabular}
\label{tab:sensor_ids}
\end{table*}

The rapid AI development in the localization domain has fundamentally transformed the landscape of positioning systems, enabling adaptability across diverse environments \cite{Ayyalasomayajula2020, Niu2019}. However, this technological progress has created a requirement for large-scale, high-quality training and benchmarking datasets that support the advancement, validation, and comparison of data-driven localization approaches \cite{Wen2015}. Data-driven methods such as fingerprinting-based approaches, neural network-based ranging, and channel state information learning, require diverse and real-world measurements to generate effective localization results across varied environmental conditions and deployment scenarios. Across indoor positioning benchmarks, ground truth rigor and metadata completeness predict localization performance more reliably than dataset size or access point density \cite{Ayub2025_Indoor_Datasets_study}.

Several indoor localization public datasets have been developed, including UJIIndoorLoc \cite{uji}, Tampere University datasets \cite{TampereU1, TampereU2}, and several other WiFi-fingerprinting, BLE RSSI, and UWB ranging datasets \cite{uwb1, wifi1, ble1}. However, these datasets suffer from limitations that restrict their utility for comprehensive algorithm development and evaluation. Notably, signal sparsity in established benchmarks can reach 93--97\%, and device-induced measurement variance spans over two orders of magnitude, posing challenges for models that aim for device-independent generalization \cite{Ayub2025_Indoor_Datasets_study}. There also exists datasets for GNSS-based localization addressing urban canyon scenarios such as UrbanNav \cite{Hsunavi.602}, and android raw GNSS measurement dataset \cite{Fu2020}. Multi-modal datasets combining radio signals with other sensor modalities such as inertial or visual data have also been proposed \cite{Abdalla2026_WiFi_CCTV_Dataset}.

The existing datasets do not explicitly address seamless indoor-outdoor transitions or mobile user scenarios. Furthermore, the most significant gap in the current localization datasets is the rarity of hybrid or multi-sensor datasets, limiting research on multi-technology fusion approaches. Hybrid and opportunistic localization remains constrained by the lack of open, multi-modal, and time-synchronized datasets that include indoor, and indoor-outdoor transitions in realistic settings.

Various radio technologies have been studied for indoor and outdoor positioning. However, these are evaluated individually and there are no datasets combining all these technologies to perform indoor-outdoor localization using signals from multiple systems. Moreover, only 22.4\% of positioning publications utilize open data and merely 7.9\% provide both open data and code \cite{ordip}, further constraining reproducibility and cross-study comparison in the field. To address this gap, this paper presents HYMN, a time-synchronized dataset encompassing five positioning technologies (GNSS, UWB, BLE, WiFi, and 5G) prepared for opportunistic radio-signal availability with explicit coverage of indoor, outdoor, and transitional environments, enabling research on opportunistic multi-technology fusion for seamless localization.

\section*{COLLECTION METHODS AND DESIGN}

\subsection{System Architecture}
The system architecture developed for HYMN data collection is set up inside an industrial hall measuring 44 m x 18 m in Torgau, Germany. The hall has a driveway with gates on each side making it perfect for indoor-outdoor positioning measurements. The architecture incorporates a combination of sensor technologies including GNSS, UWB, BLE, Wi-Fi, and 5G. These technologies when collectively used, provide an opportunistic approach facilitating effective localization.

\cref{tab:sensor_ids} gives a brief overview about the technologies used in the system architecture along with their specifications. UWB technology is utilized for its ability to provide precise distance measurements through the two-way ranging measurement principle. BLE technology is leveraged for its low power consumption and effective short-range communication. The Metirionic DMK-215 \cite{MetirionicDMK215} sensor, compatible with BLE, uses phase-based ranging. The Wi-Fi infrastructure, supported by HPE Aruba \cite{HPEArubaNetworking} access points, implements fine-time measurement (FTM) to estimate distances. Furthermore, 5G technology uses Signal-to-Noise Ratio (SNR) metric as a measurement principle to estimate positions. A Leica TS16 Total Station \cite{LeicaTS16Totalstation} is used serving the purpose of ground truth referencing of the measurement plate. This multi-sensor approach ensures a robust framework for precise data collection in an industrial hall settings.

\subsection{Measurement Setup}

\begin{figure}[ht]
\centerline{\includegraphics[width=1.0\linewidth]{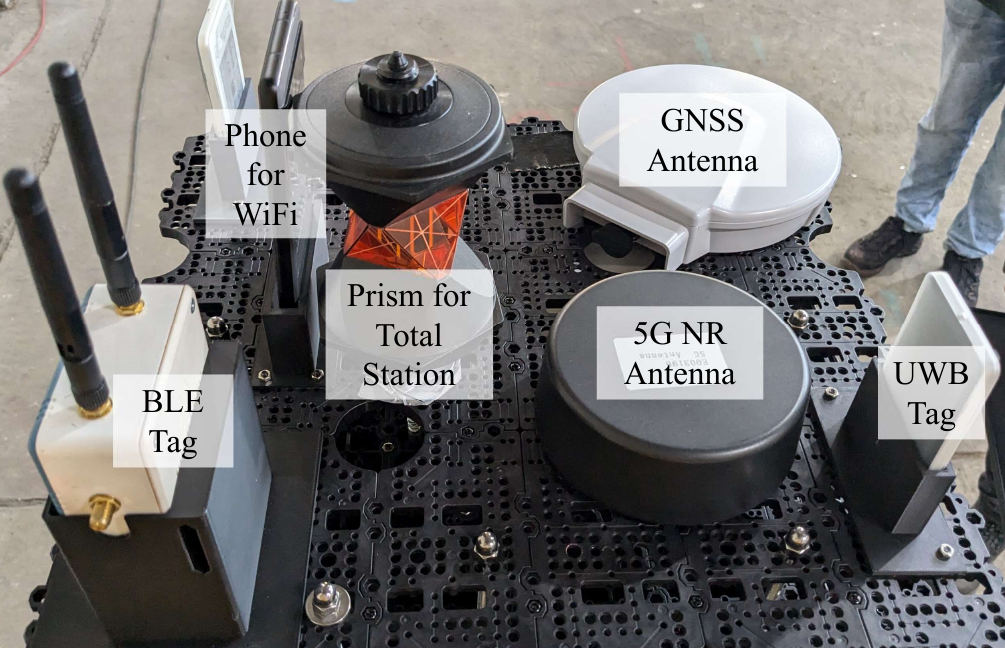}}
\caption{The measurement plate equipped with all sensors. The platform is used in static mode: it is moved manually between reference points and remains stationary with a fixed orientation for 3~min at each point while all sensors record in parallel. No data is collected during transit.}
\label{fig:measurement_plate}
\end{figure}

A mobile measurement platform was used, with all sensors mounted at fixed and known positions to ensure consistent data collection. As shown in \cref{fig:measurement_plate}, the platform included a prism for the total station, a BLE module, two UWB transceivers, a smartphone for WiFi measurements, a 5G NR antenna, and a GNSS receiver. This setup allowed simultaneous data collection across different technologies.

During measurement, sensor position references were recorded by defining the prism position as the common origin. The relative offsets are integrated into the final reference coordinates, resulting in unique, sensor-specific positions for every measurement epoch. The measurement plate was kept in a fixed orientation during the whole campaign, and this orientation was visually checked and documented at every reference point. Given the plate dimensions of approximately 30$\times$30\,cm, even a worst-case orientation error of $10^\circ$ corresponds to a positional offset on the order of a few centimetres at the outer sensors, which is well below the ranging residuals of all technologies in the dataset and therefore negligible in the overall error budget.

The anchors are distributed in the measurement hall and located at different positions to cover the whole area of measurement points. A detailed information about the geometric placement of the anchors and measurement points is available in \cite{Michler2025}. The data comprises measurement points collected at various locations along a predefined trajectory that begins outdoors, continues through the indoor area of the hall, and exits to another outdoor area. It also includes data for locations inside the hall.

\subsection{Coordinate Systems}

\begin{figure}[ht]
\centerline{\includegraphics[width=1.0\linewidth]{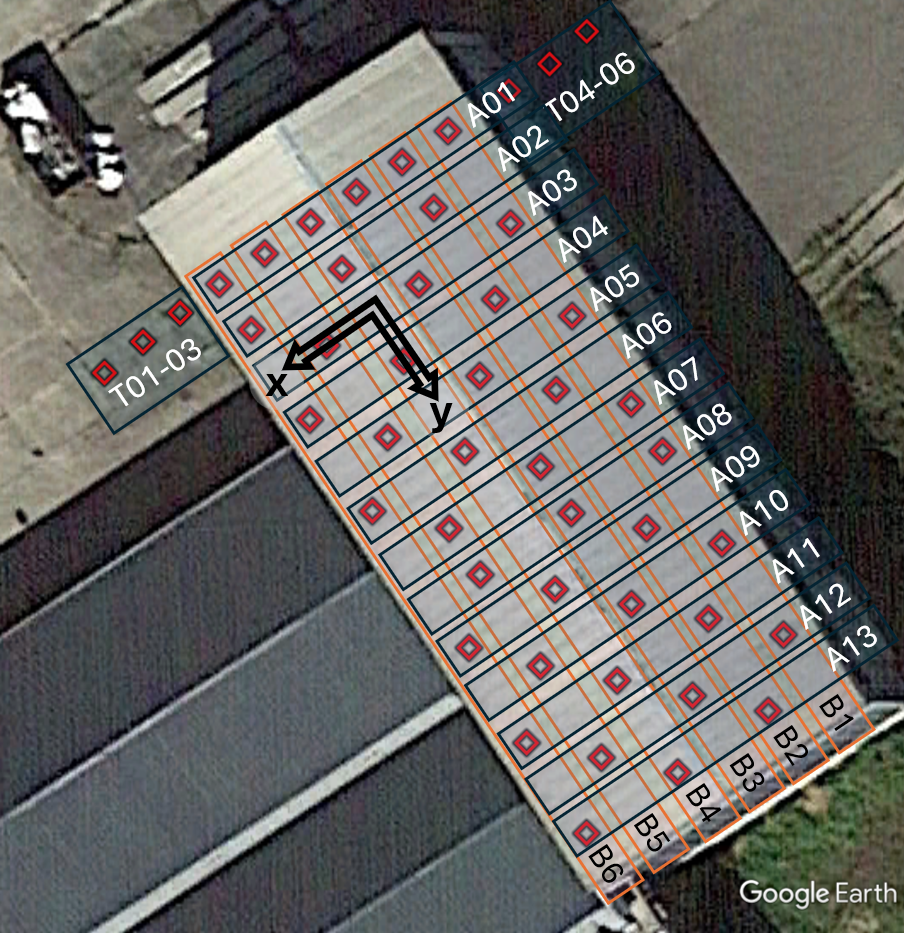}}
\caption{2D locations of measurement points inside and outside the hall shown on a geo-referenced aerial image (North-up). The arrows indicate the orientation of the local Cartesian coordinate frame. Base imagery \textcopyright 2025 Google Earth. \cite{GoogleEarth2021}}
\label{fig:measurement_locations}
\end{figure}

The spatial distribution of the measurement points is illustrated in ~\cref{fig:measurement_locations} along with the position of the reference TS used for accurate alignment. Each individual measurement point has been assigned a unique identification number for clear referencing and organization throughout the data collection and analysis process. The individual sensors are distributed across a dedicated measurement plate with fixed spatial offsets (as illustrated in \cref{fig:measurement_plate}). The transformation from local coordinates to the global UTM33N/ETRS89 frame was performed using a 2D affine transformation. The specific rotation matrix, translation vectors, and implementation details are provided in the supplementary documentation within the project repository.

\begin{table*}[ht]
    \centering
    \caption{Coordinate Reference Systems (CRS) used in the dataset.}
    \label{tab:coordinate_systems}
    \begin{tabular}{@{}lll@{}}
        \toprule
        \textbf{Name}  & \textbf{Axes / Units} & \textbf{Description} \\ \midrule
        Local cartesian coordinates & $x, y, z$ [\unit{\meter}] & Local tangent plane, needs 6-parameter affine transformation for conversion into UTM\\
        UTM33N / ETRS89  (EPSG:25833)           & $E, N, h$ [\unit{\meter}] & Projected grid for regional mapping and referencing with surveying points. \\
        ECEF / WGS84  (EPSG:4978)            & $X, Y, Z$ [\unit{\meter}] & Earth-Centered Cartesian for GNSS calculation. \\ \bottomrule
    \end{tabular}
\end{table*}

\subsection{Data Collection}
HYMN was constructed by collecting three essential elements: reference positions, measurement timestamps, and log files for each technology. At each of the 48 reference points (36 indoor, 12 transition), the platform was held stationary with a fixed orientation for 3~minutes while all sensors recorded in parallel, and no data were logged while the platform was in transit between points. Since the per-sensor update rates are not strictly deterministic (for example, UWB scheduling depends on anchor responses), the sample count varies slightly across technologies; on average, each 3-minute recording yields approximately 1\,480 BLE, 1\,407 UWB, and 1\,333 WiFi measurements, with exact per-point counts obtainable from the processed data by grouping on \texttt{point\_id}. The static acquisition mode is dictated by the ground truth reference: the Leica TS16 total station operates in static mode, so dynamic acquisitions fall outside the scope of this release and are a natural next step for a future version. The reference positions represent the true locations of the measurement plate at each measurement point, determined using the high-precision TS. By obtaining the reference position of the prism mounted on the measurement plate, and knowing that all the sensors are fixed to the same plate, the exact coordinates of each technology can be calculated by applying their respective offsets.

Measurement timestamps were recorded for every data entry to enable precise temporal alignment between different sensor systems. These timestamps help to categorize the data records by matching the log timestamps with the ground truth timestamps, allowing for accurate cross-comparison and synchronization in the subsequent processing phase.

\begin{figure}[ht]
\centerline{\includegraphics[width=1\linewidth]{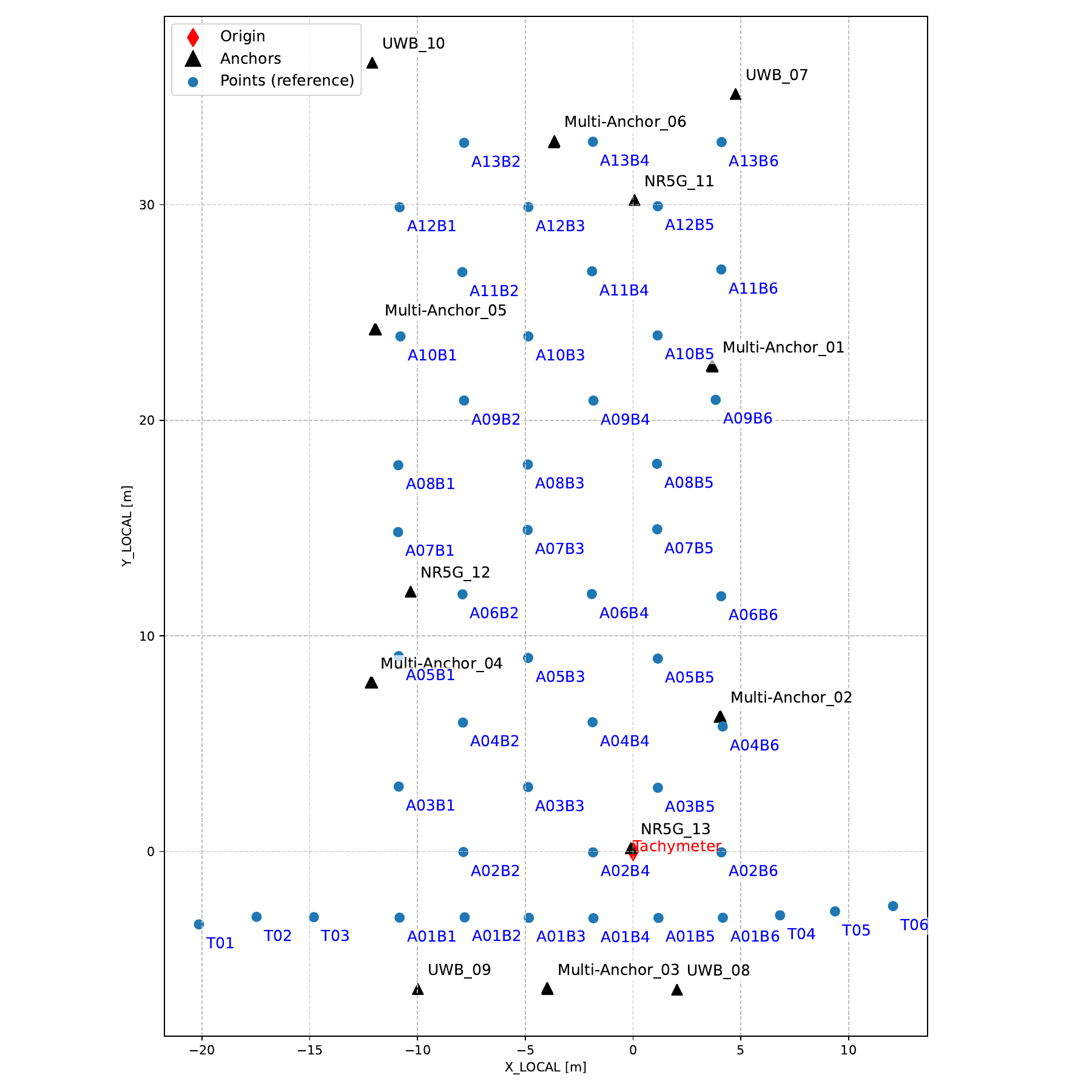}}
\caption{Layout map showing measurement grid (A1-A13, B1-B6, T1-T6) with WiFi, UWB, and BLE anchor points overlaid on test site in the local Cartesian coordinate frame. Note that the axis orientation differs from \cref{fig:measurement_locations} as the two figures represent different coordinate reference systems (see \cref{tab:coordinate_systems}).}
\label{fig:sat_image}
\end{figure}

For each technology, specific logging scripts were implemented to collect the raw logs containing the complete set of recorded measurements. These log files include the sensor-specific observables/readings necessary for subsequent data processing and analysis. The raw log data refers to each location with different identifier as shown in \cref{fig:sat_image}. These are later formatted in data processing stage where each measurement location is assigned a systematic point id based on structured grid. The indoor area uses a two-dimensional grid with numbering along the longer side of the hall (column) from A1 - A13 and along the shorter side of the hall (row) from B1 - B6. The first digits in the \texttt{point\_id} show the grid column, and the rest indicate the row. Measurement points with a T prefix mark reference locations outside the main hall and act as transition or boundary markers.

\subsection{Data Processing}
Creating a unified dataset from the five different technologies requires a systematic data processing workflow to ensure compatibility and readability for further analysis tasks such as localization, filtering, model training, and performance evaluation. Since each sensor's logging format is distinct, the processing pipeline varies across individual sensors.

The final processed dataset adheres to a unified format, enabling seamless integration and multi-sensor analysis.

\begin{figure}[ht]
\centerline{\includegraphics[width=1.0\linewidth]{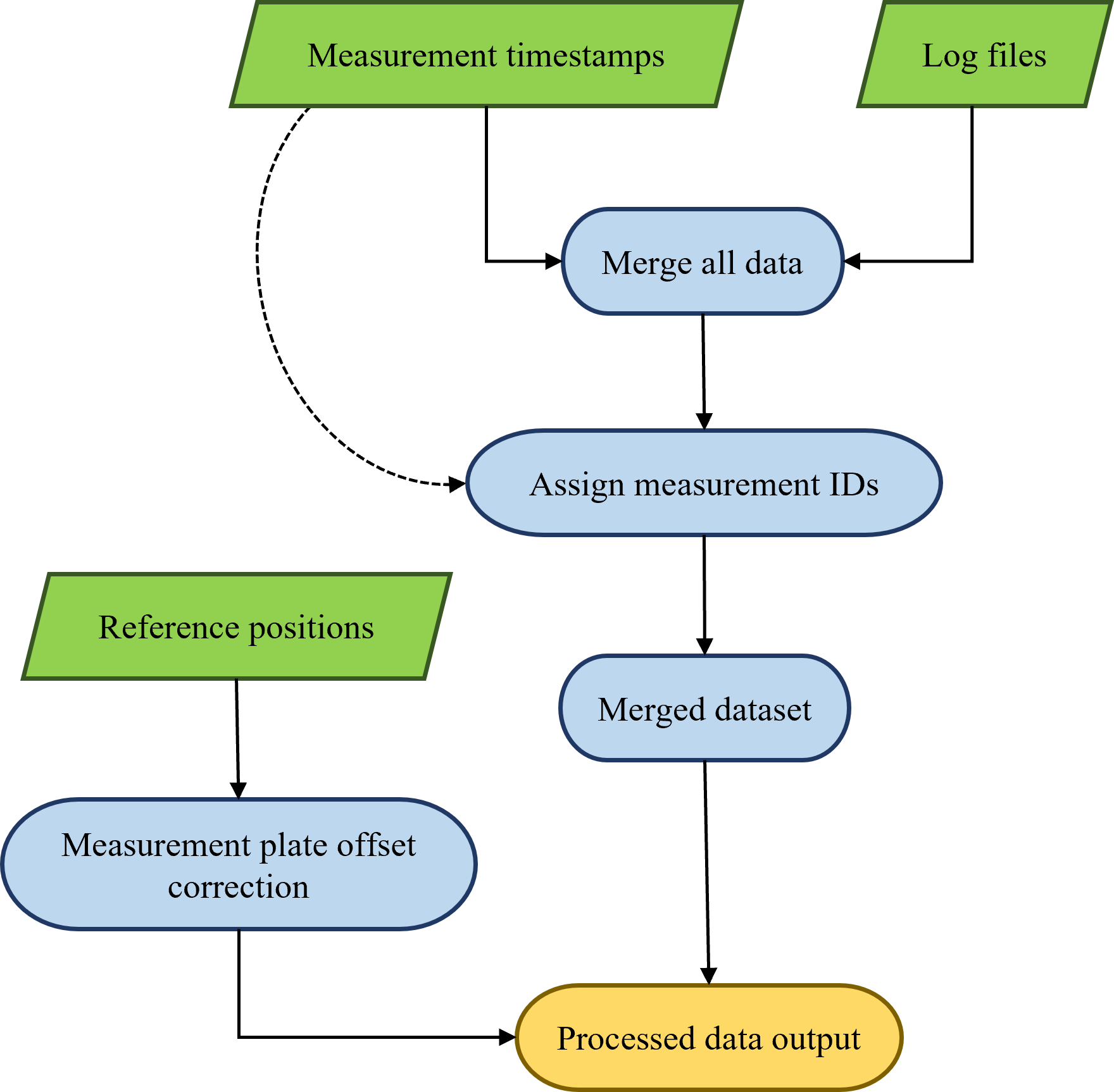}}
\caption{Processing workflow from measurements to processed output.}
\label{fig:workflow}
\end{figure}

A brief workflow is illustrated in \cref{fig:workflow} which outlines the sequential steps used to transform raw measurement data into a multi-sensor calibrated dataset suitable for analysis. In the workflow diagram, green boxes represent the input data sources and the blue boxes shows the data processing steps. The process begins with measurement timestamps and log files from all five technologies, which are merged to create a comprehensive dataset, segmented by temporal boundaries, containing all the recorded measurements.

Specific to GNSS, the data is preprocessed to include satellite positions, following the general scheme of providing known anchor locations within every measurement epoch. These satellite positions are not computed in real time at the point of data collection; instead, they are derived during the pre-processing stage. This is done using the \texttt{gnss\_lib\_py} Python package \cite{knowles_glp_2024}, which fetches the precise ephemeris products published after the collection period and uses them to calculate satellite coordinates for the corresponding measurement epochs.

After merging the raw measurements from all sensor subsystems, the combined records are sorted by their individual timestamps and every observation is assigned its own unique \texttt{measurement\_id} for traceability. The \texttt{measurement\_id} therefore identifies a single raw reading and does not group observations into temporal epochs; the reference point at which a reading was taken is identified separately through the \texttt{point\_id} field, which links every row to its static ground truth position.

As the ground truth refers to the position of the prism mounted on the measurement plate rather than the individual receivers, it is necessary to determine both the relative and absolute positions of each sensor's receiver on the plate. This is achieved by applying known geometric offsets between the prism and the receivers. These offsets enable the transformation of the prism's coordinates into precise positions for every sensing technology present on the plate.

Furthermore, the 5G positioning system used for data collection operated in a local coordinate reference frame with a different orientation from the other sensing systems. To ensure spatial consistency across all datasets, additional processing steps were required, including coordinates system alignment and transformation. This involved rotating and translating the 5G coordinate data into the common reference frame. In certain cases, the UWB observables from multiple anchors are logged out of sequence due to varying reception orders. To correct this inconsistency, the range readings are re-ordered to maintain a unified anchor sequence, ensuring that all sensor logs follow an identical structure format. In this context, re-ordering means sorting the range measurements per epoch so that they always follow a strict sequence based on anchor ID.

In the next step, we join the ground truth coordinates processed initially with the corresponding measurement IDs to include the receiver's coordinates with each recorded measurement data across all the sensors. This merging step is essential for combining positional ground truth with the respective sensor readings for our analysis and validation steps. Finally, to complete the standardization process, all columns are renamed according to a unified naming convention, also shown in a sample data format in \cref{tab:measurement_table1} and \cref{tab:measurement_table2}. This uniform labeling eliminates ambiguity, ensures consistency, and prepares the dataset for further analysis without sensor-specific adjustments.

Sensor synchronization is handled at two levels. At acquisition time, every logging device obtains its timestamps from a common NTP reference via the public internet pool, giving a shared wall-clock reference at the millisecond level that is sufficient for correct temporal ordering at the per-sensor update rates of roughly 1\,Hz, without providing ranging-level time synchronization. Since all points were recorded statically with a stationary total-station reference, no post-hoc clock alignment was required. In the processed dataset the sensors remain asynchronous: each measurement keeps its individual timestamp and is placed on a common temporal axis (see \cref{tab:sensor_merge}) rather than being resampled to a fixed grid. This preserves the raw timing of every event and keeps the dataset directly compatible with sequential filters that consume multi-rate, asynchronous observations, as is standard in hybrid radio positioning~\cite{Guo_gnss_uwb_time_calibration_2023}. \cref{tab:sensor_merge} illustrates this structure as a conceptual example and does not represent actual data entries.

\begin{table}[ht]
    \centering
    \caption{Exemplary structure of merged multi-sensor dataset. Values are illustrative and do not correspond to actual dataset indices. Each column encodes the index of the associated observation in the respective subsystem table (WiFi, BLE, UWB, GNSS, 5G NR), while ‘–’ indicates absence of a measurement at the given timestamp.}
    \begin{tabular}{lcccccc}
        \toprule
        \textbf{ts [Unix time, ms]} & \textbf{point\_id} & \textbf{wifi} & \textbf{ble} & \textbf{uwb} & \textbf{gnss} & \textbf{nr5g} \\
        \midrule
        1729686520045.0 & A13B6 & -  & 0 & 0  & -  & -  \\
        1729686521631.0 & A13B6 & -  & 1 & -  & 0  & -  \\
        1729686522834.0 & A13B6 & 0 & 2  & -  & 1  & 0  \\
        1729686524032.0 & A12B6 & -  & 3 & 1  & -  & -  \\
        1729686525651.0 & A12B6 & 1 & 4  & -  & 2  & -  \\
        \bottomrule
    \end{tabular}
    \label{tab:sensor_merge}
\end{table}

\section*{VALIDATION AND QUALITY}

Dataset quality assurance followed established best practices for open research data in indoor positioning, as systematically reviewed by Anagnostopoulos et al.~\cite{ordip}. The dataset is published in Zenodo with persistent DOI assignment \cite{HYMN_Dataset_zenodo_2025}, ensuring long-term accessibility and citability. Comprehensive metadata documentation addresses critical attributes frequently undefined in existing datasets: precise environment characterization, explicit ground truth collection methodology, and technical specifications for all sensor systems. All processing scripts for data loading and coordinate transformations are provided under the MIT license, supporting transparent reuse and extension by the research community.

The technical quality of the dataset was validated through comparison with high-precision ground truth measurements obtained using a Leica TS16 total station. All anchor positions and measurement points were surveyed with millimeter-level accuracy, establishing the reference framework for ranging residual analysis.

Ranging accuracy was assessed by computing residuals between measured ranges and true geometric distances derived from surveyed positions. Table~\ref{tab:statistic} summarizes the statistical characteristics of ranging errors for each terrestrial positioning technology. These characteristics are derived from substantial sample sizes for each technology, with approximately 49,700 BLE, 62,700 UWB, and 63,100 WiFi ranging measurements, as well as approximately 284,000 5G SNR observations and 103,500 GNSS pseudoranges.

\begin{figure}[ht]
\centerline{\includegraphics[width=1\linewidth]{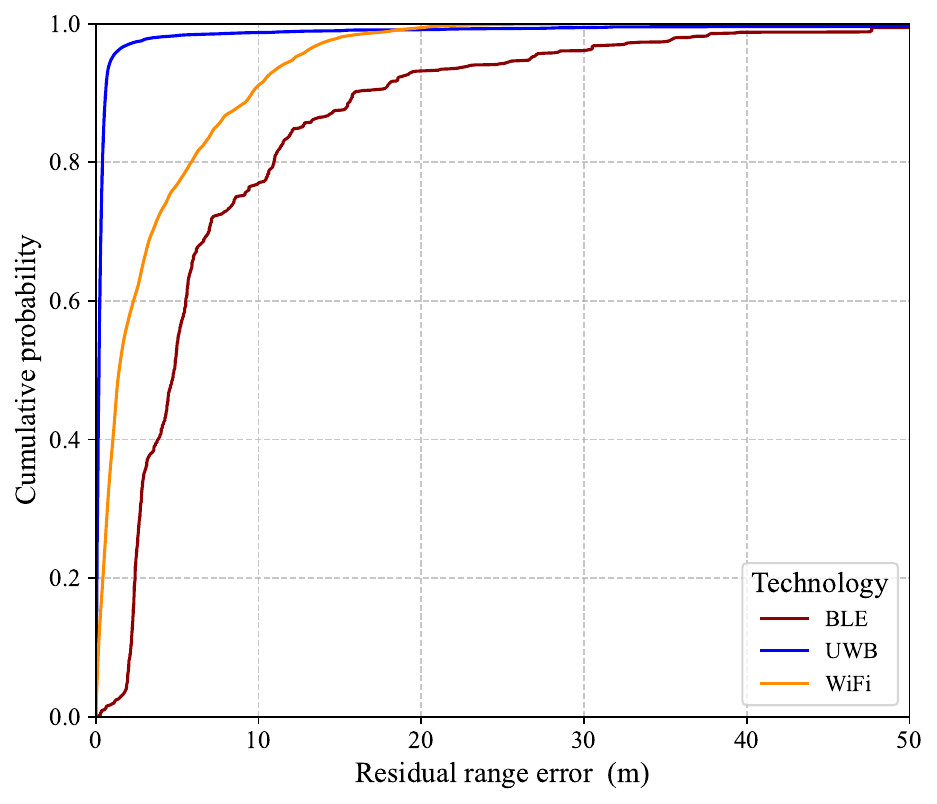}}
\caption{Cumulative probability of absolute ranging residuals for UWB, WiFi, and BLE.}
\label{fig:ecdf}
\end{figure}

UWB demonstrates the highest accuracy with a mean residual of 1.03\,m and median of 0.20\,m, while BLE exhibits the largest systematic bias (mean: 8.07\,m) and higher variability (standard deviation: 12.93\,m). WiFi achieves intermediate performance with mean residual of 3.30\,m. The empirical cumulative distribution functions (Figure~\ref{fig:ecdf}) reveal distinct error characteristics: UWB shows rapid convergence with 90\% of residuals below 2\,m. However, its 99th percentile (15.18\,m) confirms the presence of occasional significant outliers, contributing to its elevated standard deviation. BLE and WiFi, in contrast, exhibit longer tails indicative of multipath effects and measurement outliers.

Specifically, 95\% of BLE residuals are below 26.74\,m (with 99th percentile at 47.68\,m), and for WiFi, 95\% are below 12.20\,m (with 99th percentile at 18.27\,m).

\begin{table*}[!ht]
\centering
\caption{Statistical summary of ranging residual errors for BLE, UWB, and WiFi.}
\label{tab:statistic}
\begin{tabular}{@{}lcccccccc@{}}
\toprule
\textbf{Tech.} & \textbf{Count} & \textbf{Mean (m)} & \textbf{Median(m)} & \textbf{75\% (m)} & \textbf{95\% (m)} & \textbf{99\% (m)} & \textbf{Std. (m)} \\
\midrule
BLE  & 49745 & 8.07 & 4.81 & 8.64 & 26.74 & 47.68 & 12.93 \\
UWB  & 62745 & 1.03 & 0.20 & 0.36 & 1.00 & 15.18 & 9.00 \\
WiFi & 63121 & 3.30 & 1.44 & 4.48 & 12.20 & 18.27 & 4.11 \\
\bottomrule
\end{tabular}
\end{table*}

GNSS measurements were validated using between-satellite single differencing (BSSD), yielding an empirical standard deviation of 6.8\,m for code-based pseudoranges across all measurement zones. A comprehensive analysis of system-specific performance, including availability statistics, bias characterization, and environmental dependencies, is provided in \cite{Michler2025}.

\section*{RECORDS AND STORAGE}
\begin{figure}[ht]
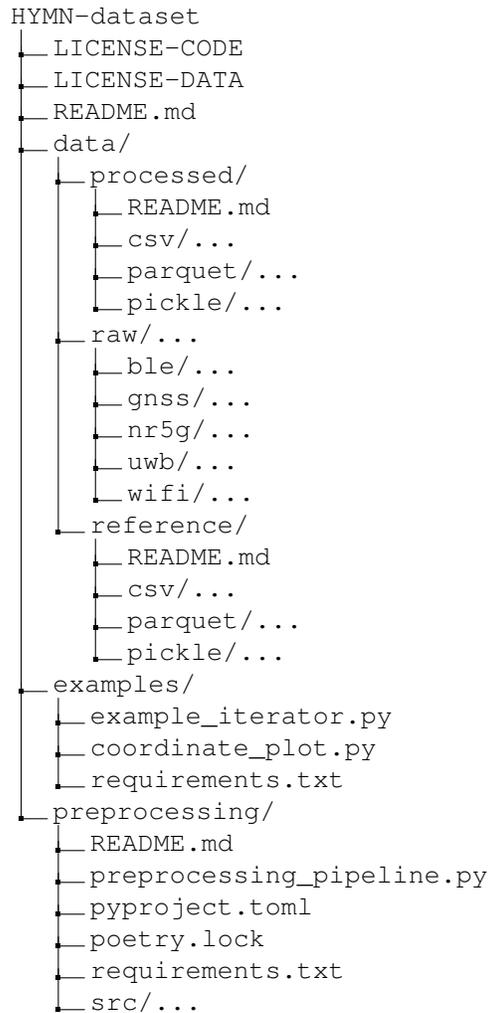

    \centering
    \begin{minipage}{0.8\textwidth} 
    \dirtree{%
        .1 HYMN-dataset.
        .2 LICENSE-CODE.
        .2 LICENSE-DATA.
        .2 README.md.
        .2 data/.
        .3 processed/.
        .4 README.md.
        .4 csv/....
        .4 parquet/....
        .4 pickle/....
        .3 raw/....
        .4 ble/....
        .4 gnss/....
        .4 nr5g/....
        .4 uwb/....
        .4 wifi/....
        .3 reference/.
        .4 README.md.
        .4 csv/....
        .4 parquet/....
        .4 pickle/....
        .2 examples/.
        .3 example\_iterator.py.
        .3 coordinate\_plot.py.
        .3 requirements.txt.
        .2 preprocessing/.
        .3 README.md.
        .3 preprocessing\_pipeline.py.
        .3 pyproject.toml.
        .3 poetry.lock.
        .3 requirements.txt.
        .3 src/....
    }
    \end{minipage}
    \caption{Hierarchical directory structure of the repository.}
    \label{fig:structure}
\end{figure}

The HYMN dataset is provided alongside comprehensive preprocessing scripts, organized as illustrated in \cref{fig:structure}. The repository is structured to support efficient storage, access, and processing of multi-sensor data across five distinct technologies. The raw data is also available which includes the time-stamped records for all the technologies.

The \texttt{data/} directory serves as the central hub for all dataset components:
\begin{itemize}
    \item \textbf{Processed Data:} Measurements are provided in three standardized formats (\texttt{csv}, \texttt{parquet}, and \texttt{pickle}) to suit different analytical requirements. This includes the merged dataset, which aligns measurements from multiple technologies on a unified timeline.
    \item \textbf{Raw Data:} Original, unprocessed timestamped records for each technology, including BLE, GNSS, NR5G, UWB, and WiFi. These files are released as-is to preserve the full preprocessing chain and allow independent reconstruction of the processed dataset. As a consequence, the raw logs contain occasional empty rows or incomplete records that originate from restarted measurement sessions or from transient radio-link failures typical of real-world wireless systems. Such artefacts are filtered out during preprocessing and do not appear in the processed dataset.
    \item \textbf{Reference Data:} Reference points and time reference used to align and calibrate measurements, ensuring spatial and temporal consistency across the processed datasets.
\end{itemize}

\subsection*{Multi-technology Merged Data Alignment}
The merged data concept acts (\texttt{data/.../merged}) as a synchronization layer that aligns asynchronous sensor events from multiple technologies onto a unified temporal axis. By providing a merged list of timestamps mapped to specific row indices in system-specific data tables, it serves as a lightweight lookup table that avoids data redundancy while maintaining cross-references. To iterate through this dataset, a user can loop over the timestamps in the merged table and use the corresponding indices as foreign keys to instantly retrieve high-resolution hardware logs from the respective subsystem files. This process is illustrated in \texttt{examples/example\_iterator.py}.

\subsection*{Preprocessing and Utility Scripts}
The \texttt{preprocessing\/} folder contains the logic required to transform raw sensor data into final formats. The \texttt{src\/} subfolder houses core Python source files for cleaning and transforming data, while \texttt{preprocessing\_pipeline.py} streamlines the entire workflow.

To facilitate ease of use, the \texttt{examples/} directory provides utility scripts \texttt{example\_iterator.py} and \texttt{coordinate\_plot.py}, which demonstrate how to interact with and visualize the dataset. Documentation and dependency requirements (e.g., \texttt{README.md}, \texttt{requirements.txt}, and \texttt{pyproject.toml}) are distributed throughout the repository to ensure full reproducibility of the processing environment.

\begin{table*}[ht]
\centering
\caption{Example data table for UWB, BLE, and WiFi systems}

\begin{tabular}{@{}lllll@{}}
\toprule
\textbf{point\_id} & \textbf{ts}            & \textbf{anchor\_ids}               & \textbf{ranges}                             & \textbf{[X,Y,Z]\_LOCAL\_BLE}                \\ \midrule
A13B6        & 1729686520045 & "['BLE1', 'BLE2', 'BLE3', 'BLE4', 'BLE5']" & "{[}52.9, 38.05, 46.39, 23.34, 61.32{]}" & "(4.019, 32.81, 3.284)" \\
A04B6        & 1729686520958 & "['BLE1', 'BLE2', 'BLE3', 'BLE4', 'BLE5']" & "{[}43.03, nan, 39.73, 29.56, nan{]}"    & "(4.07, 5.705, 3.264)" \\ \bottomrule
\label{tab:measurement_table1}

\end{tabular}
\end{table*}

\begin{table*}[ht]
\centering
\caption{Example data table for 5GNR system}

\begin{tabular}{@{}llllll@{}}
\toprule
\textbf{point\_id} & \textbf{ts}            & \textbf{anchor\_ids}             & \textbf{pos}          & \textbf{SNR}                             & \textbf{[X,Y,Z]\_LOCAL\_NR5G}\\ \midrule
A12B5       & 1729686795005& "{[}'NR5G\_11', 'NR5G\_12', 'NR5G\_13'{]}" & "(-6.219, 32.349)" & "{[}34.64, 31.46, 31.38{]}" & "(1.26, 29.873, 3.345)" \\
A06B4        & 1729692056551 & "{[}'NR5G\_11', 'NR5G\_12', 'NR5G\_13'{]}" & "(-3.819, 11.96)" & "{[}25.14, 27.75, 26.25{]}"    & "(-1.797, 11.885, 3.344)" \\ \bottomrule
\label{tab:measurement_table2}

\end{tabular}
\end{table*}

The \textit{preprocessing} folder contains the files to prepare and organize raw sensor data. The \textit{src} folder contains the Python source files applied for data processing, including functions for cleaning, transforming, and standardizing the sensor readings. The \textit{preprocessing\_pipeline.py} file streamlines the workflow by using the scripts in the \textit{src} folder across the data, efficiently producing datasets ready for further analysis.

\subsection*{File Descriptions}
As file formats for processed data, CSV was selected to ensure long-term accessibility and straightforward usability as suggested in \cite{ordip}. In addition, the Parquet file format was chosen to provide a language-agnostic and efficiently parsable columnar standard, while Pickle was included to enable direct integration into Python-based processing workflows.
To ensure consistency across the terrestrial radio-based systems UWB, BLE, 5G-NR and WiFi, a standardized tabular structure is used, as illustrated in \cref{tab:measurement_table1}. Each row within this table represents a unique measurement and contains the following information:

\begin{itemize}
\item \texttt{point\_id} - Unique measurement point identifier.
\item \texttt{ts} - Timestamp of the measurement.
\item \texttt{anchor\_ids} - Identifiers of the reference anchors.
\item \texttt{ranges} - Measured ranges between the tag/device and the corresponding anchors.
\item \texttt{X\_local, Y\_local, Z\_local} - Reference position of sensor/receiver.
\end{itemize}

The 5G system dataset has a similar format of storage but it differs slightly, as it provides signal-to-noise ratios (SNR) instead of range measurements and position estimates as well. \cref{tab:measurement_table2} summarizes the format in which the 5G dataset is stored. The additional information provided includes:

\begin{itemize}
\item \texttt{pos} - Provided 5G position in local coordinate frame \texttt{(x,y)} in meters, as estimated by the proprietary localization.
\item \texttt{SNR} - SNR values measured between the receiver and the corresponding anchors.
\end{itemize}

In contrast to terrestrial systems, the GNSS data is stored as more complex structure due to the nature of satellite-based observations. It includes raw measurements (e.g., pseudorange, carrier phase, and Doppler) with satellite state information (e.g., positions, velocities, and clock parameters). Additionally, RINEX files are provided in the raw data folder. Despite the difference in schema, all files are consistently aligned through the common identifier \texttt{point\_id}, which provide a unified reference across all the technologies.

\section*{INSIGHTS AND NOTES}
HYMN comprises heterogeneous multi-technology measurements intended for the development, analysis, and validation of seamless indoor–outdoor localization approaches. The utilized technologies are standardized and widely adopted, underscoring the dataset’s relevance and transferability to real-world applications. Its distinct contribution compared to existing datasets lies in the explicit inclusion of an indoor–outdoor transition component and the comprehensive coverage of multiple radio-based positioning technologies, thereby enabling opportunistic multi-sensor data fusion. By providing complete metadata documentation, open preprocessing code, and persistent DOI assignment, HYMN directly addresses the documented deficits in open science practices within the positioning community \cite{ordip, Ayub2025_Indoor_Datasets_study}.

Several technical considerations are important when working with this dataset. The positioning systems employed operate under different network topologies: while 5G, UWB, and BLE rely on infrastructure-based processing, GNSS and WiFi perform device-based positioning. This architectural difference poses challenges for synchronized data collection, as each system's measurement timing depends on different processing pipelines and communication protocols. Consequently, temporal alignment across systems required careful timestamp management during data acquisition, with all measurements referenced to a common time base to enable meaningful cross-technology fusion analysis.

Researchers should also note that coordinate frame alignment varies across technologies. While GNSS operates in a global reference frame, terrestrial systems utilize local coordinate frames. The preprocessing pipeline addresses this by transforming all measurements into a unified local coordinate system, with transformation parameters documented in the reference files. Users developing fusion algorithms must account for these coordinate system differences, particularly when integrating global and local positioning modalities. Particular value is provided for research on localization using mobile consumer devices, such as smartphones, as contemporary models support all included technologies.

Consequently, HYMN can serve as a benchmark for validating multi-sensor, on-device localization algorithms. Future data acquisitions may be conducted directly using smartphones to incorporate device-specific hardware and antenna characteristics, thereby further enhancing representativeness and practical applicability.

\section*{SOURCE CODE AND SCRIPTS}

The HYMN dataset and all associated code are openly available on GitHub (\url{https://github.com/TUD-ITVS/HYMN-dataset}) with persistent identifier \href{https://doi.org/10.5281/zenodo.17979434}{DOI: 10.5281/zenodo.17979434}. The repository includes:

\begin{itemize}
\item \textbf{Processed data}: Multi-format datasets (CSV, Parquet, Pickle) for all technologies, including merged multi-sensor and minimal timestamp availability files
\item \textbf{Raw measurement data}: Unprocessed sensor logs with timestamps and technology identifiers
\item \textbf{Preprocessing pipeline}: Complete Python implementation (\texttt{preprocessing\_pipeline.py}) with modular source scripts for data cleaning, transformation, coordinate alignment, and standardization
\item \textbf{Reference data}: Ground truth positions from total station measurements, sensor offset calibrations, and coordinate transformation parameters
\item \textbf{Documentation}: README with dataset description, file formats, usage examples, and citation information
\end{itemize}

All preprocessing scripts are implemented in compatibility with Python 3.8+ and utilize standard scientific libraries including NumPy, Pandas, and the \texttt{gnss\_lib\_py} \cite{knowles_glp_2024} for GNSS data processing. As of the current release, the Python version needs to be prior to 3.13 due to compatibility of the upstream \texttt{gnss\_lib\_py} dependency. The codebase is released under an open-source license to facilitate reproducibility and community contributions.

\section*{ACKNOWLEDGEMENTS AND INTERESTS}

\textbf{Conceptualization:} AMi, AMu, PS, JN, HU, OM.
\textbf{Data Curation:} AMi, AMu.
\textbf{Formal Analysis:} AMi, AMu, PS.
\textbf{Funding Acquisition:} PS, OM.
\textbf{Investigation:} AMi, AMu, PS, JN, HU, OM.
\textbf{Methodology:} AMi, AMu, PS, JN, HU.
\textbf{Project Administration:} PS, OM.
\textbf{Resources:} OM.
\textbf{Software:} AMi, AMu, PS, JN, HU.
\textbf{Supervision:} PS, OM.
\textbf{Validation:} AMi, AMu, PS, JN, HU.
\textbf{Visualization:} AMi, AMu, JN.
\textbf{Writing~-- Original Draft:} AMi, AMu, PS.
\textbf{Writing~-- Review \& Editing:} AMi, AMu, PS.

\medskip
All authors reviewed the manuscript. The article authors have declared no conflicts of interest.

\medskip
\noindent\begin{tabular}{m{5.5cm} m{2.5cm}}
This work has been partially funded by the German Federal Ministry for Digital and Transport (BMDV) within the project IDEA (FKZ: 19OI22020C). We thank MRK Management Consultants GmbH for facility access. & \includegraphics[width=1\linewidth]{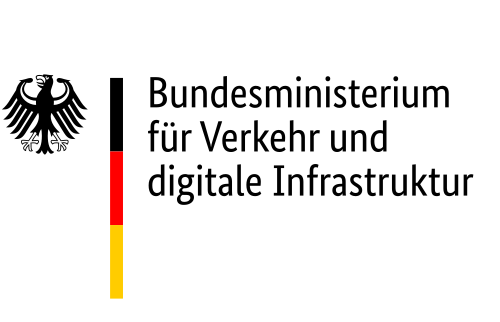} \\
\end{tabular}

\bibliographystyle{IEEEtran}
\bibliography{lit.bib}
\balance

\end{document}